\documentclass[aps,amssymb,pra,twocolumn,showpacs,floatfix,showkeys]{revtex4}

\usepackage{epsf}
\usepackage{amsmath}
\usepackage{graphicx}

\begin{document}

\title {Acoustically driven x-ray emission and matter collapse in lead}

\author{F. Fern\'andez$^{1}$, A. M. Loske$^{1}$, and B. I. Ivlev$^{2}$}

\affiliation
{$^{1}$Centro de F\'{\i}sica Aplicada y Tecnolog\'{\i}a Avanzada, Universidad Nacional Aut\'onoma de M\'exico, Blvd. Juriquilla 3001, 76230 Qu\'{e}retaro, QRO, M\'exico\\
and\\
$^{2}$Instituto de F\'{\i}sica, Universidad Aut\'onoma de San Luis Potos\'{\i}, 78000
San Luis Potos\'{\i}, SLP, M\'exico\\}


\begin{abstract}

The action of focused underwater weak shock waves on a lead sample was revealed to be not restricted by a mechanical influence only. A strong unexpected x-ray emission was registered from the lead 
foil exposed to those shock waves ({\it sound into x-rays}) which were extremely adiabatic compared to processes of x-ray generation. The lead foil, exposed to shock waves, lost a part of its area 
having the shape of a polygonal hole of the size of $\sim 2mm$. The missing polygon of lead foil looks as a delicately removed part with no damage at the hole surroundings as it should be after a  
mechanical breaking. This points to a non-mechanical mechanism of hole formation. That missing polygonal lead matter seems to be ``disappeared'' because the total lead volume was reduced by that 
amount after exposure to acoustic waves ({\it matter collapse}). Both paradoxical phenomena cannot be explained by a combination of known effects and a fundamentally new mechanism is required to 
underlie them. The concept of electron anomalous states, which encouraged the experiments and specified main features of them, is likely that mechanism.

\end{abstract} \vskip 1.0cm



\maketitle


\section{INTRODUCTION}
\label{intr}
External perturbations of matter may result in x-ray radiation. It can occur at braking of moving electrons (Bremsstrahlung) \cite{BET,HAU,LANDAU,CHE} and also under electron transitions to lower 
energy levels in atoms (characteristic radiation) \cite{BON,SHA,VAN}. Known x-ray emission phenomena \cite{MAT,DUN,HUA,PUTT,HOR} result from those basic ones. 

Acoustic action on matter can be strong. Shock waves in liquids result in cavitation \cite{COL,VOG,LAU,PEC,BEN}. Ultrasonically driven bubbles in liquids may emit bursts of light when they 
collapse (sonoluminescence) \cite{PEC1,HIL,PUT1,PUT2,BRE}. Shock waves in a soft tissue can provide a strong mechanical action \cite{LOSKE}. In solids acoustic energy may be converted into visible 
light \cite{OST} or infrared radiation \cite{AHN} (acoustoluminescence). 

X-ray emission can occur under action on solids of extremely strong shock waves, with the pressure of $200GPa$ \cite{GIZ}. In this case external mechanical forces on solids are comparable with the 
atomic ones. The intensity of those shock waves is four orders of magnitude larger than in the experiments reported here (weak shock waves). Furthermore, x-ray emission from radiative shock waves 
in astrophysics can occur due to heating of a matter up to high temperature \cite{NYM}. In our experiments weak shock waves result in the negligible increase of temperature, $\sim 0.01K$.

In this article we study completely different phenomena compared to the ones mentioned above. Acoustic pulses, propagating through lead, resulted in x-ray emission. This phenomenon may seem to be 
just one more on the list of various types of x-ray emission. This impression is not correct, since the x-ray emission revealed here cannot be explained by a combination of known effects. 
In addition, it is also accompanied by a paradoxical matter collapse which has not been observed before. It turns out that a fundamentally new mechanism has to underlie those two phenomena. It is 
likely that the concept of electron anomalous states \cite{IVLEV}, which encouraged the experiments and specified main features of them, is this mechanism. Here we only introduce the discovered 
phenomena. A detailed study of them is out of the scope of this article.

We exposed a ``sandwich'' (see details in Sec.~\ref{exp}), containing a lead foil, to underwater shock waves. The passage of the acoustic pulse through a particular point in the lead occurs during 
approximately $\Delta t\simeq 10^{-10}s$. A perturbation with this characteristic time cannot excite electrons up to the $keV$ energy scale, corresponding to the frequency $\omega\sim 10^{18}s^{-1}$. 
The formal probability of one quantum excitation up to the $keV$ energy is of the type $\exp(-\omega\Delta t)$ \cite{IVLEV1,IVLEV2}. In a multiquanta absorption, $\omega\Delta t\sim 10^{8}$ quanta 
should be involved resulting in the probability of the type above \cite{IVLEV1,IVLEV2}. Therefore, a characteristic x-ray emission in the $keV$ region is impossible. Also there are no conditions in 
lead for Bremsstrahlung of that energy since conduction electrons adiabatically follow the acoustic wave acquiring its velocity $\sim 10^5cm/s$. 
\begin{figure}[h!]
\centering
\includegraphics[width=7cm , bb=5 5 350 330]{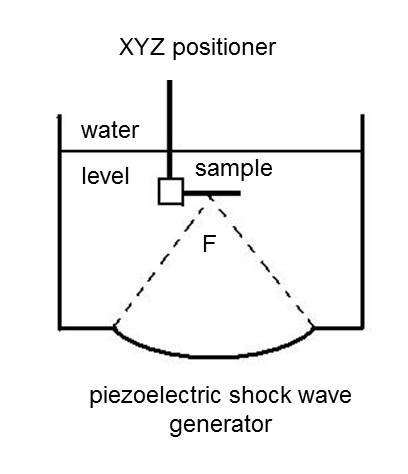}
\caption{\label{fig1}Sketch of the experimental arrangement. The piezoelectric transducer, cross-sectioned in the figure, has a spherical shape. Pressure waves are emitted from the transducer at a 
rate of $0.5Hz$ and focused on the center $F$ of the arrangement inside a water tank.}
\end{figure}

Slow varying acoustic perturbation, in principle, can produce a spatial charge separation resulting in an electric field locally increasing an electron energy with a possibility of a
subsequent quanta emission. For example, such mechanism resulted in acoustoluminescence in a semiconductor due to a lattice defect separation \cite{OST}. But in the bulk lead fast electron 
processes (keeping a fixed Fermi level) with a characteristic time $10^{-15}s$ \cite{ABR} equalize any acoustically caused charge separation. The only possibility for a charge separation is a 
lead - dielectric border in the sandwich used in our experiments. One can suppose a deformation of surfaces of those solids when they locally loose contact forming a cavity with opposite charges 
on separated surfaces. This would remind the charge separation at peeling of an adhesive tape when the separated charges resulted in glow discharge and x-ray Bremsstrahlung \cite{PUTT}. But in our 
experiment such discharge avalanche cannot be formed due to air which is unavoidable between the two surfaces. The hypothetical cavity would be filled out by air at atmospheric pressure. In 
this case an accelerating electron would be braked by air molecules which is not sufficient for x-ray Bremsstrahlung.

One can conclude that a conversion of the acoustic energy into $keV$ x-rays by known mechanisms is impossible. However, we observed a strong x-ray emission when shock waves come from water and 
reflect off a sample containing the lead foil ({\it sound into x-rays}). The energy of x-ray quanta is estimated to be in the $keV$ range. The source of the observed x-rays was not in the water 
(including cavitation bubble dynamics), since it was screened by a relatively thick copper film. As argued in this paper, the source was the lead. 
\begin{figure}[h!]
\centering
\includegraphics[width=5cm , bb=5 5 350 350]{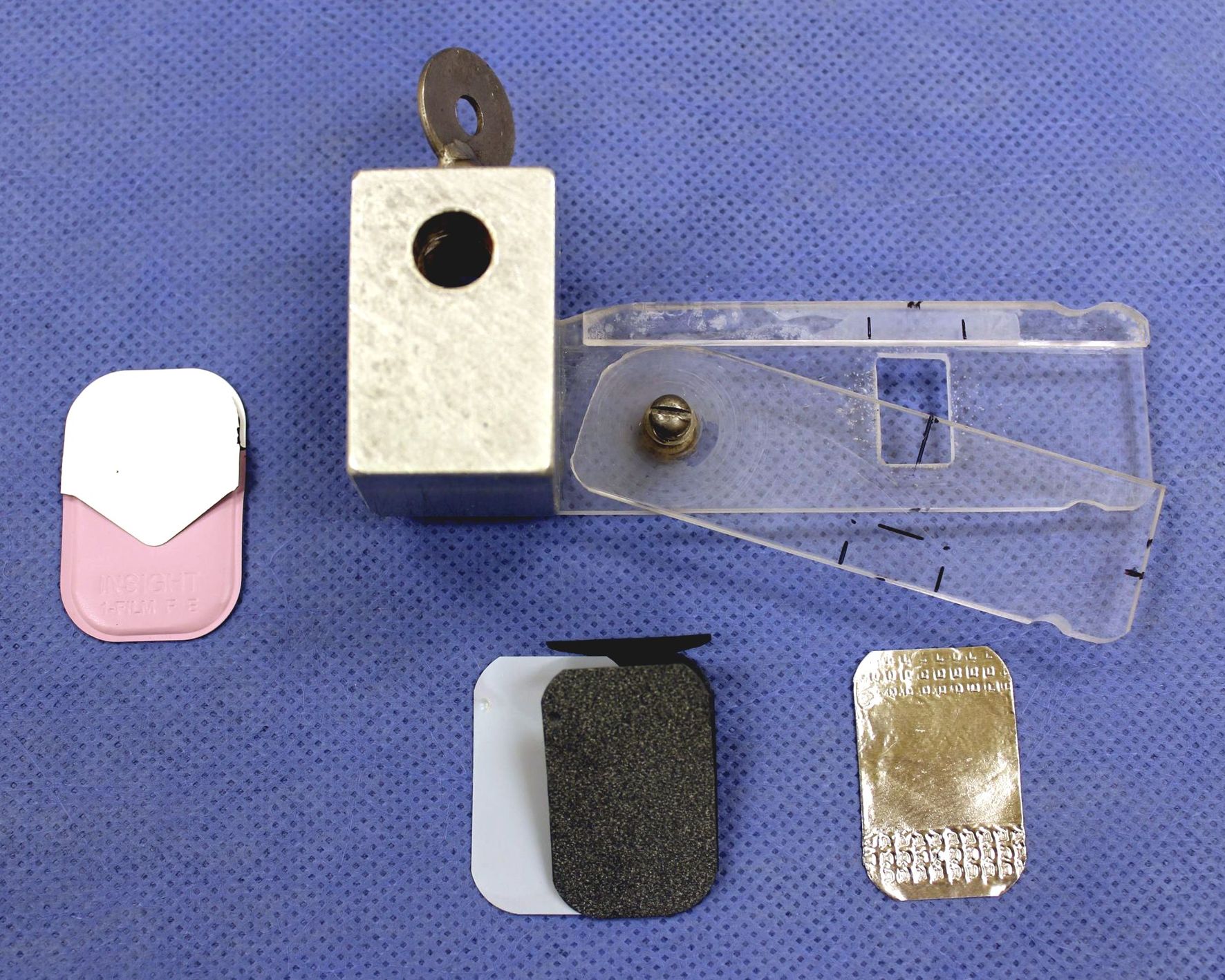}
\caption{\label{fig2}Disassembled sample holder. The plastic envelope, the x-ray film with the light protecting paper, and the lead foil are shown separately (not as a sandwich). The 
sandwich oppression occurs outside the $10\times 15mm^2$ window in the rigid plastic plate.}
\end{figure}

The other experimental observation is also unusual. Under shock wave exposure a small area of the lead foil (of $2mm$ size and of polygonal shape) ``disappears'' because the total lead volume was 
reduced by that amount after exposure to acoustic waves ({\it matter collapse}). Moreover, the missing polygon of lead foil looked as a delicately removed part with no damage at the hole 
surrounding as it should at a mechanical breaking. This points to a non-mechanical mechanism of hole formation. A resemblance between two phenomena is manifested in our experiments. 

\section{MATERIALS AND METHODS}
\label{exp}
Shock waves, as an acoustic type of matter perturbation, are widely used in various fields \cite{LOSKE}. These waves consist of a narrow front and a relatively long release wave \cite{BEN}.
Usual applications of shock waves relate to their mechanical actions on matter \cite{BEI,PHI,ILO,KLA}. Shock waves in water can result in cavitation arising from nucleation sites and 
microbubbles \cite{PHI,ILO,KLA}. During an individual bubble collapse, high-speed microjets of water may be generated \cite{LAU}. These microjets have sufficient energy to puncture thin plastic or 
metal film after a multiple action. 

The main element of the setup used was a Piezolith 2501-based shock wave source (Richard Wolf GmbH, Knittlingen, Germany). It consists of approximately 3000 small piezoelectric elements, arranged 
on a bowl-shaped (radius $345mm$) aluminum backing (Fig.~\ref{fig1}) with an aperture of $474mm$. The electric circuit basically consists of capacitor charging unit and a discharge control system. 
The piezoelectric elements are insulated from water by a flexible membrane. A high voltage discharge across the array of piezoelectric elements generates an abrupt expansion of all elements, 
producing pressure waves that travel towards the center $F$ of the device. The superposition of propagating pressure waves, due to a hydrodynamic nonlinear distortion, produces a shock wave in the 
vicinity of the focal region (see Fig.~\ref{fig1}). In this region, in the shape of a cigar measuring about $17\times 3mm$, aligned along the axis of symmetry, the positive pressure pulse has more 
\begin{figure}[h!]
\centering
\includegraphics[width=4.2cm , bb=5 5 300 350]{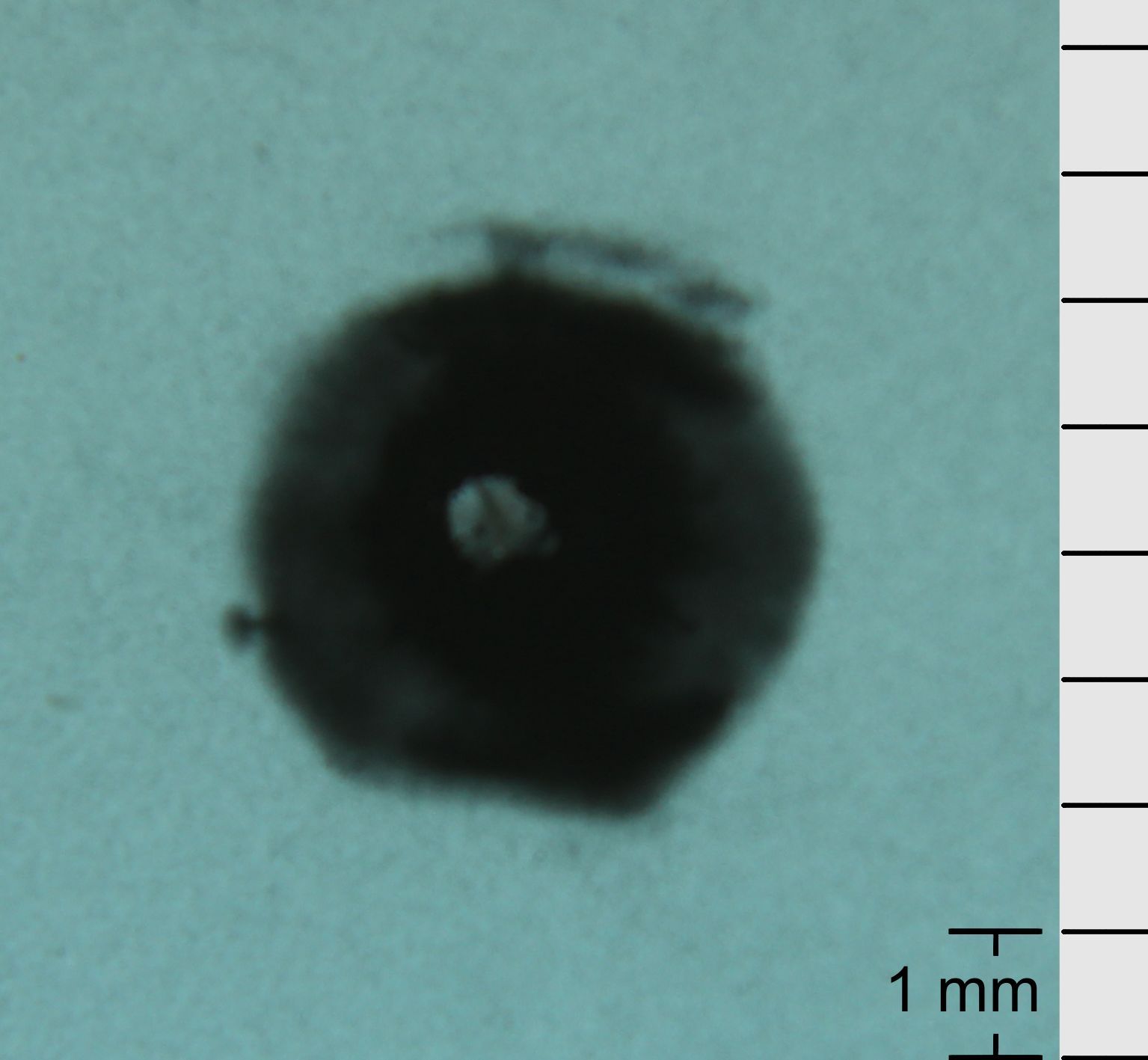}
\caption{\label{fig3}Image on the x-ray film after exposure to 1000 shock waves with a peak positive pressure amplitude of $26MPa$. (Shock wave) - (plastic envelope) - ($0.4mm$ thick copper foil) 
- (x-ray film) - (lead foil) - (plastic envelope).}
\end{figure}
\begin{figure}[h!]
\centering
\includegraphics[width=5cm , bb=5 5 350 350]{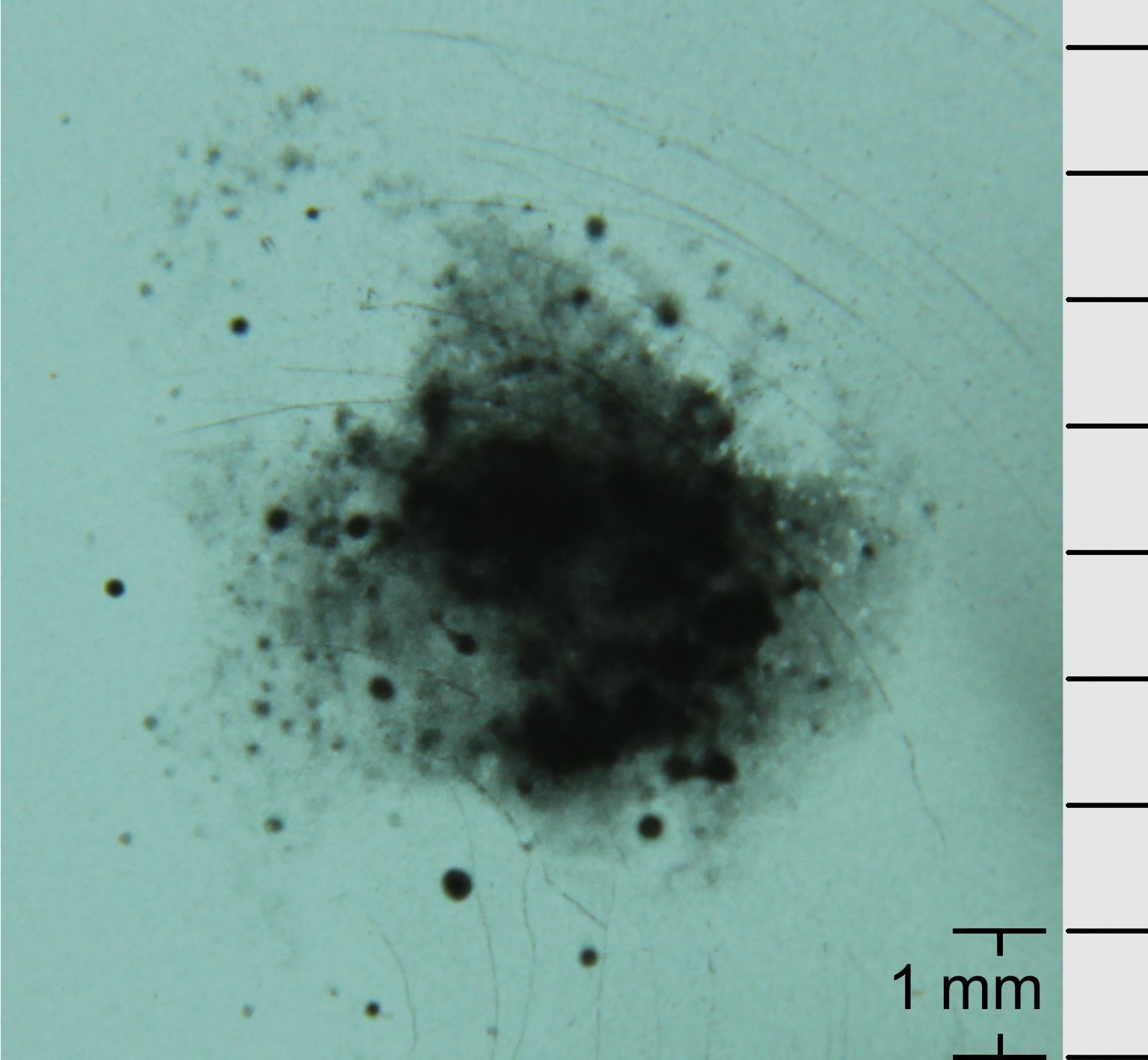}
\caption{\label{fig4}Image on the x-ray film after exposure to 250 shock waves with a peak positive pressure amplitude of $26MPa$. (Shock wave) - ($0.3mm$ thick plastic film) - (x-ray film) - 
(lead foil).}
\end{figure}
an $50\%$ of the maximum peak positive amplitude.

\begin{figure}[h!]
\centering
\includegraphics[width=5cm , bb=5 5 350 350]{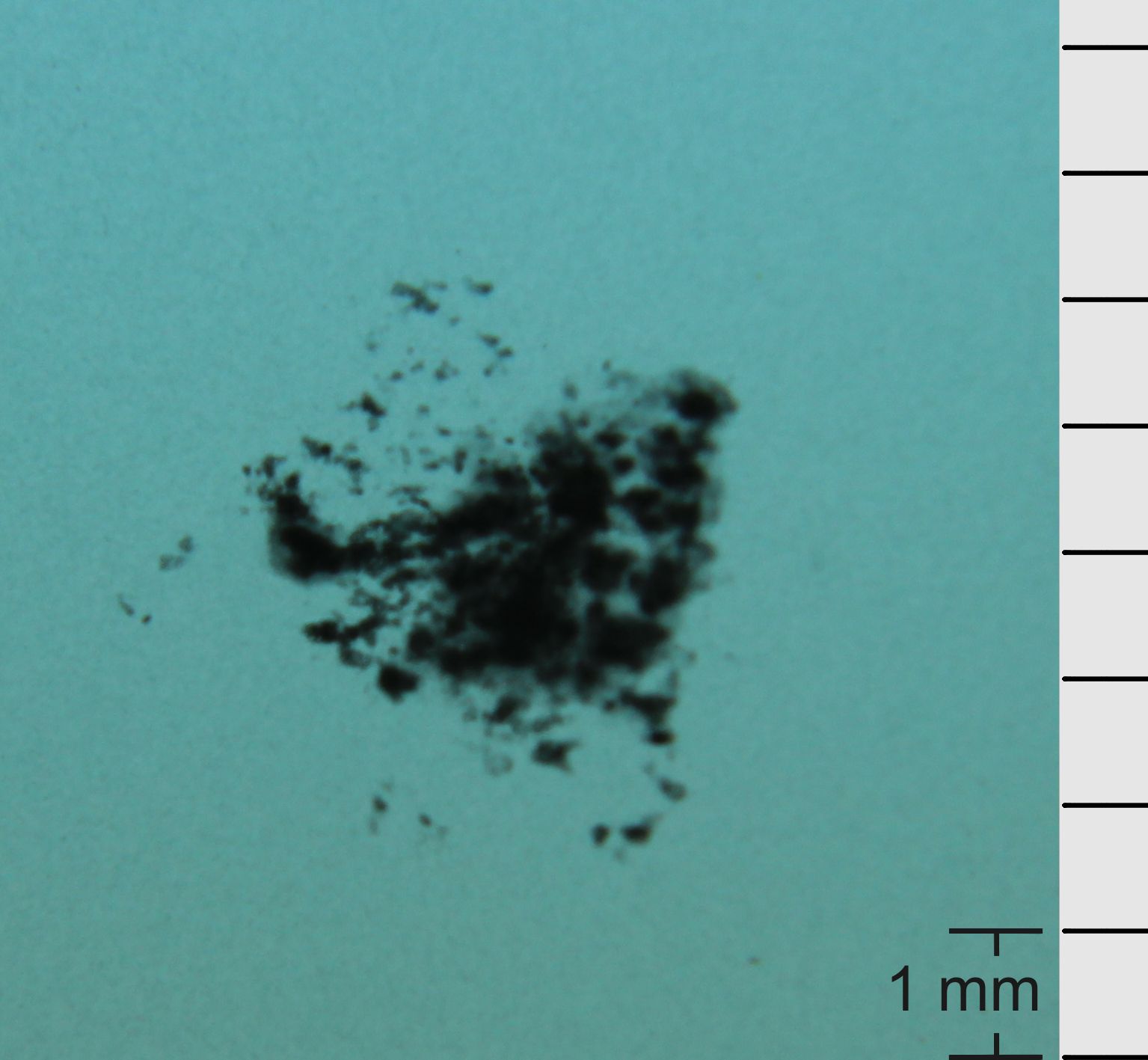}
\caption{\label{fig5}Image on the x-ray film after exposure to 250 shock waves with a peak positive pressure amplitude of $29MPa$. (Shock wave) - (plastic envelope) - (black paper) - (x-ray film) 
- (black paper) - (lead foil) - (plastic envelope).}
\end{figure}

\begin{figure}[h!]
\centering
\includegraphics[width=5cm , bb=5 5 350 350]{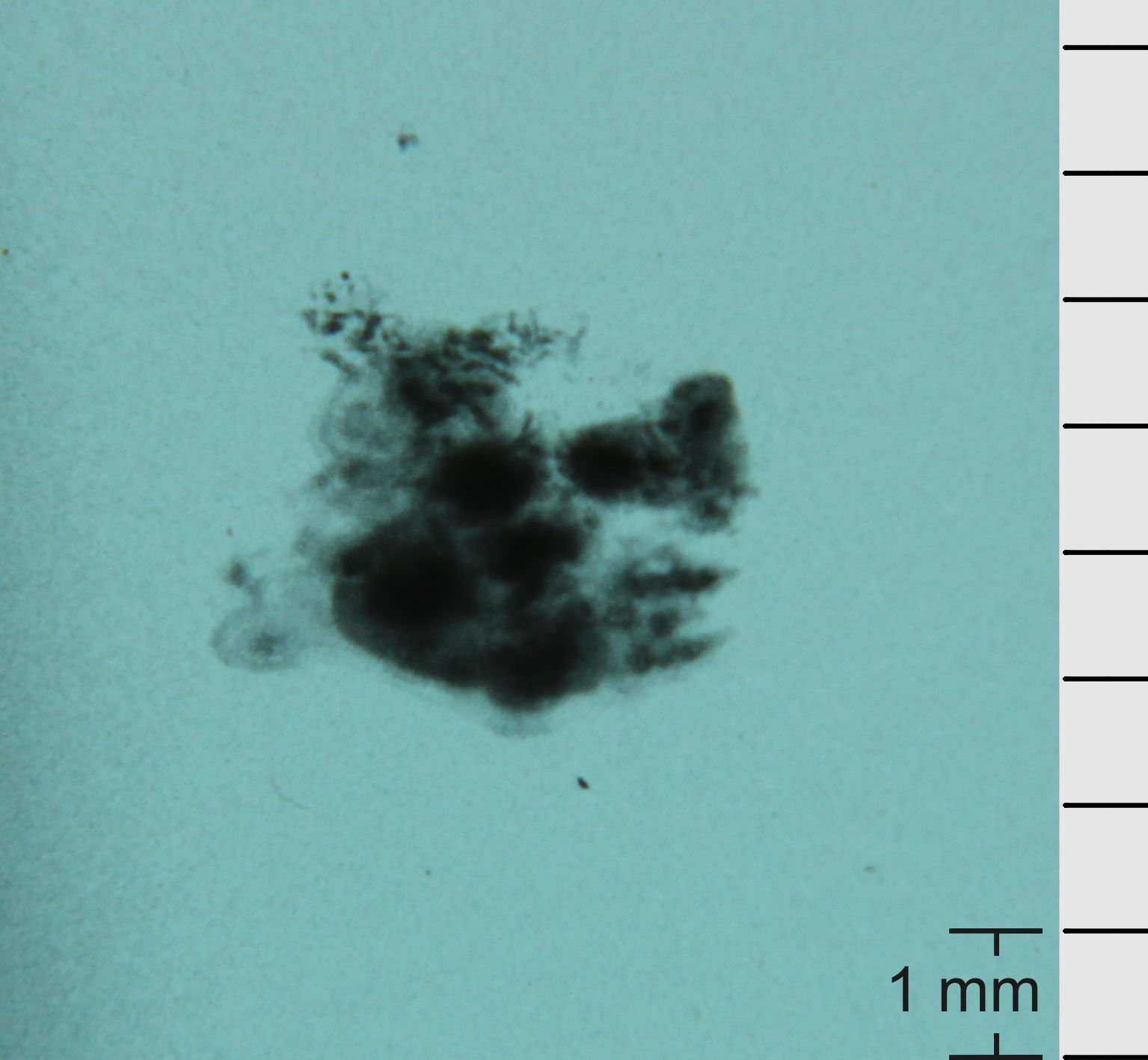}
\caption{\label{fig6} Image on the x-ray film after exposure to 250 shock waves with a peak positive pressure amplitude of $29.5MPa$. (Shock wave) - (plastic envelope) -(lead foil) - (black paper) 
- (x-ray film) - (black paper) - (plastic envelope).}
\end{figure}

\begin{figure}[h!]
\centering
\includegraphics[width=6cm , bb=5 5 400 400]{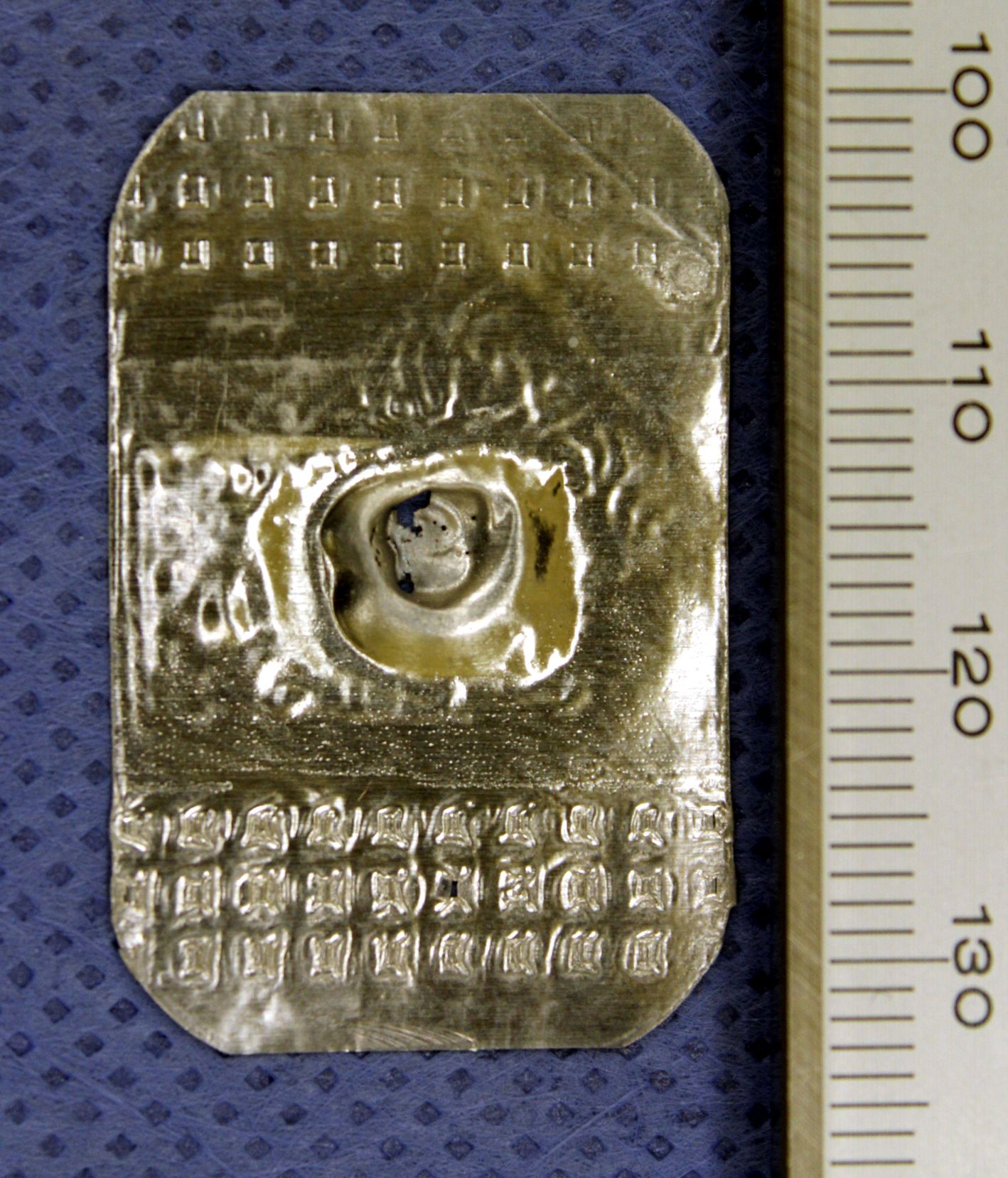}
\caption{\label{fig7}Lead foil after exposure to 130 shock waves with a peak positive pressure amplitude of $29.3MPa$. The foil was immersed in water at the focal plane with the water level $40mm$ 
above $F$. The shock wave velocity, calculated by Eq.~(\ref{0}), was $v\simeq 1.523\times 10^5cm/s$.}
\end{figure}

\begin{figure}[h!]
\centering
\includegraphics[width=6cm , bb=5 5 400 400]{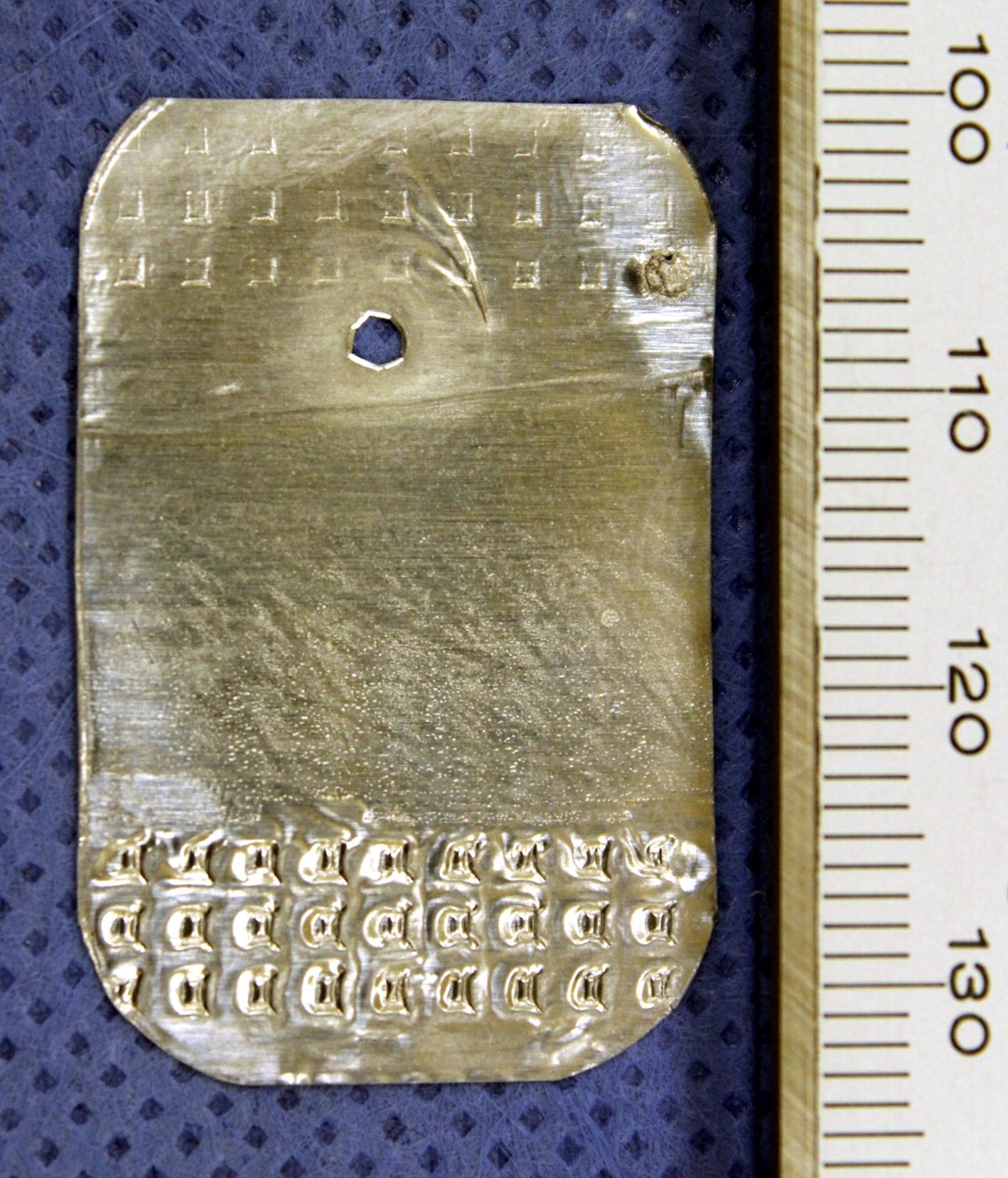}
\caption{\label{fig8}Comparison to Fig.~\ref{fig7}. The lead foil was inside the plastic envelope and exposed to 250 shock waves with a peak positive pressure amplitude of $29MPa$. The shock wave 
velocity, calculated by Eq.~(\ref{0}), was $v\simeq 1.523\times 10^5cm/s$. Figs.~\ref{fig8} - \ref{fig14} correspond to the sequence (shock wave) - (plastic envelope) - (lead foil) - (x-ray film)
- (plastic envelope).}
\end{figure}

\begin{figure}[h!]
\centering
\includegraphics[width=5cm , bb=5 5 350 350]{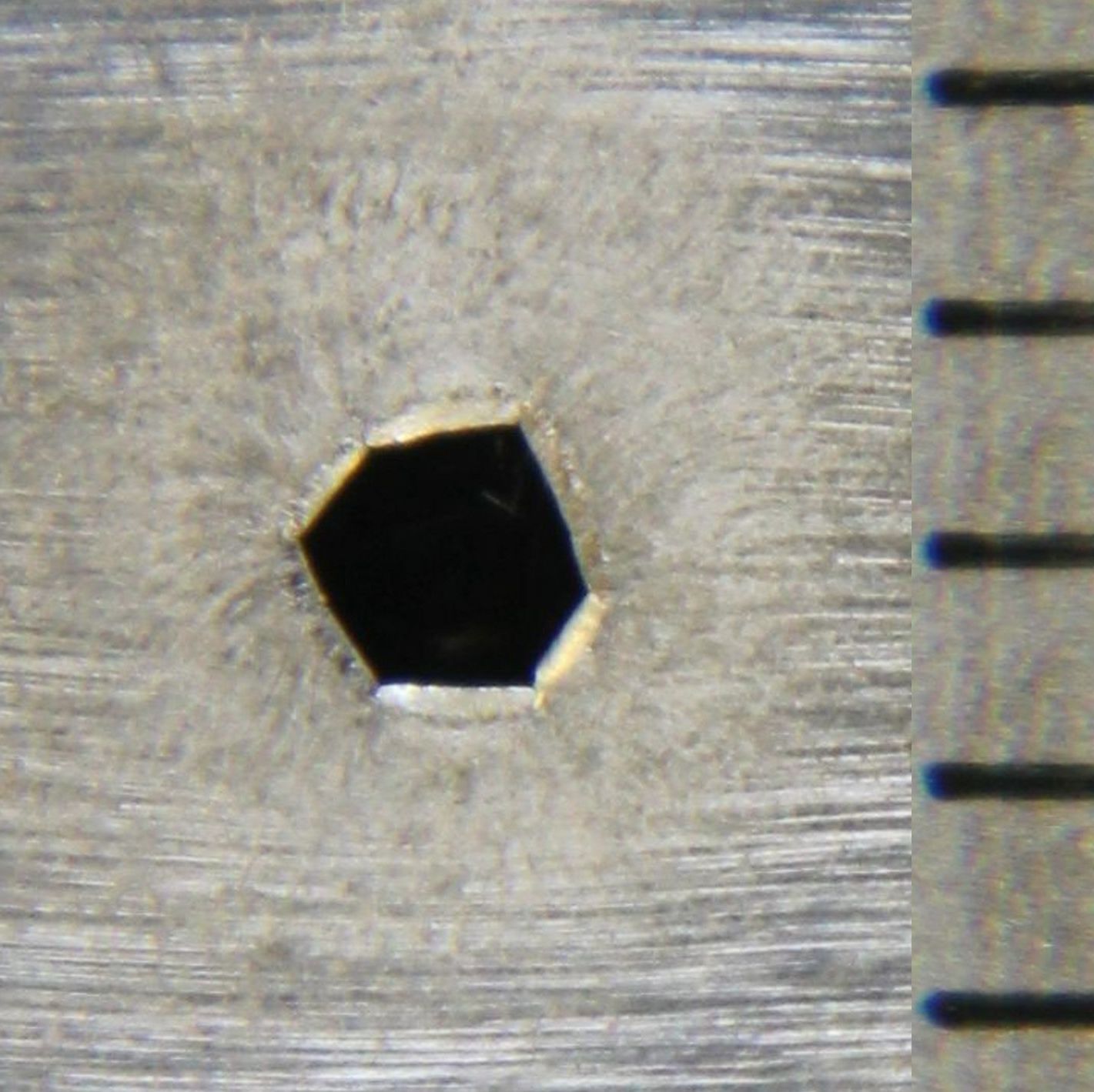}
\caption{\label{fig9}Hexagonal hole in the lead foil after exposure to 250 shock waves with a peak positive pressure amplitude of $26MPa$. The shock wave velocity, calculated by Eq.~(\ref{0}), was 
$v\simeq 1.519\times 10^5cm/s$. The scale is in millimeters.}
\end{figure}

All samples were exposed to pressure waves generated at a voltage between 2.5 and $5kV$ and a discharge rate of $0.5Hz$. At this voltage, the pressure waveform, measured at $F$ with a PVDF needle 
hydrophone (Imotec GmbH, W\"{u}rselen, Germany), had a peak positive amplitude at the pressure interval $(15 - 33)MPa$. Pressure values are specified in Figs.~\ref{fig3} - \ref{fig11}. The 
temperature inside the water tank was approximately $20^oC$. Special care was taken to avoid air gaps or bubbles at the water-sample interface or between layers inside the sample. The shock wave 
velocity $v$ in water is connected to the pressure $p$ in the front by the relation \cite{COL,PEC}
\begin{equation}
\label{0}
p=275\bigg\{\left[1.526\left(\frac{v}{s_0}-1\right)\right]^{2.310}-1\bigg\}MPa,
\end{equation}
where $s_0=1.48\times 10^5cm/s$ is the sound velocity in water at $20^oC$. 

As a sample in Fig.~\ref{fig1} a ``sandwich'', based on Carestream Dental-Insight intraoral x-ray film ($22\times 35mm^2$) and manufactured by Carestream Health Inc. (Rochester NY, USA), was 
used. Each individual film was packaged in a light-tight plastic (vinyl) envelope to protect the film from light. The front side is supposed to face the x-ray tube; the back side has an opening tab 
to open the film during processing. A black paper wrapper ($0.10mm$ thick) protects the film base (sheet of cellulose acetate of $0.20mm$ thickness) from light and damage. The x-ray sensitive 
emulsion faces towards the front side of the envelope. A lead foil ($0.06mm$ thick) is on the back side. 

Each plastic envelope, fixed with the specially designed Lucite holder shown in Fig.~\ref{fig2}, was fastened horizontally and centered at the focus $F$. The x-ray sensitive emulsion was facing 
towards the shock waves coming from the water. Waves propagated through the outer plastic cover, the front black paper, the x-ray film, the black paper, the lead foil, and the back plastic cover. 
The water level was set $40mm$ above $F$. A small strip of water resistant duck tape was used to seal the opening tab of each x-ray film envelope. 

Shock waves, created near the focus $F$ were not the only mechanical effect on the sample. Another mechanical effect was due to cavitation in the focal region. This occurs because the water 
contains nucleation sites and microbubbles \cite{PHI,ILO,KLA}. Cavitation produces secondary shock waves which also impact the sample. In addition, shock wave-induced microjets of water are 
generated during individual or multiple bubble collapses \cite{LAU}. These microjets have sufficient energy to puncture thin plastic or metal film after a multiple action.

\section{SOUND INTO X-RAYS}
\label{xray}
Various types of sandwiches (samples in Fig.~\ref{fig1}) were exposed to shock waves in water. We have started the research obtaining the image shown in Fig.~\ref{fig3}, where the sandwich used 
contained a $0.4mm$ thick cooper foil. 

The exposed spot on the x-ray film in Fig.~\ref{fig3} could be interpreted as produced by x-rays emitted from the focal region in the water. The diameter of that region ($\sim 3mm$) coincides with 
the spot size; however, in that case x-rays had to be of at least $\sim 50keV$ to overcome the attenuation in the copper foil. This finding also looks strange since the spot in Fig.~\ref{fig3} has 
the obvious tendency for polygonal formation but the focal region does not have such shape. The definite exclusion of that scenario followed from the absence of an exposed spot (a clean x-ray film 
after development) after removing the lead foil from the sandwich keeping the rest as it was. 

The presence of the copper layer in the sandwich, described in Fig.~\ref{fig3}, was not crucial for the phenomenon. In the sandwich, described in Fig.~\ref{fig4}, the copper film was substituted by 
a layer of plastic but the x-ray film was still exposed. 

Visible light did not contribute to film exposure. This was proved by using light protecting black papers in the sandwich described in Fig.~\ref{fig5}. 

In the sandwich described in Fig.~\ref{fig6}, the light protected x-ray film was placed behind the lead foil. This arrangement opened the possibility to study a propagation of the emitted quanta in 
space and to measure their spectrum. 

So the source of x-rays, detected as shown in Figs.~\ref{fig3} - \ref{fig6}, was the shock wave-exposed lead foil. The underwater shock wave consists of a narrow wave front having a width 
$\Delta x\sim 0.15\mu m$ and a release wave, $\sim 1cm$, following the front (see Ref.~\cite{LOSKE} and references therein). The shortest characteristic time $\Delta t$ of acoustic perturbations 
in lead is determined by the duration of the shock wave front in water. $\Delta t$ is approximately the ratio of the width $\Delta x$ of the shock wave front and the velocity $v$ of the shock wave
\begin{equation}
\label{1}
\Delta t\simeq\frac{0.15\mu m}{1.52\times 10^{5}cm/s}\simeq 10^{-10}s.
\end{equation}

We did not measure the energy of emitted x-ray quanta precisely. However, we know that the x-ray emission reduces approximately twice after penetration through the black paper which is the usual 
light protection of the x-ray films used. It was also demonstrated that the emitted x-rays hardly penetrate through the wet black paper. Therefore, one can roughly estimate the energy of emitted 
x-ray quanta to be in $1keV$ region. This corresponds to the quanta frequency $\omega\sim 10^{18}s^{-1}$. 

In Fig.~\ref{fig6} the total time of exposure to x-rays was $250\Delta t\sim 10^{-8}s$. This produces the same darkness of the spot as an x-ray ramp from a dental x-ray tube with a $0.1s$ duration.
So the estimated x-ray flux, during the acoustic pulse action, was approximately $10^7$ times higher than a flux from a dental x-ray tube.

Which is the mechanism producing the observed x-ray emission from lead (sound into x-rays)? The perturbation with the characteristic time (\ref{1}) cannot excite electrons up to the $keV$ energy 
scale, corresponding to the frequency $\omega\sim 10^{18}s^{-1}$. The formal probability of one quantum excitation up to the $keV$ energy is of the type $\exp(-\omega\Delta t)$ \cite{IVLEV1,IVLEV2}. 
In the multiquanta absorption, $N\simeq \omega\Delta t\sim 10^{8}$ quanta should be absorbed to increase an electron energy up to the $keV$ scale \cite{IVLEV1,IVLEV2}. But in our experiment the 
acoustic perturbation of the lead is not very strong (almost elastic regime) since it relates to a variation of approximately $20\%$ of the lead density at the position of the acoustic peak. 

Therefore, the $N$-th order of a perturbation theory, a product of $N$ small values, corresponds to the probability mentioned above \cite{IVLEV1,IVLEV2}. Therefore a characteristic x-ray emission 
in the $keV$ region is impossible. Also there are no conditions in lead for Bremsstrahlung of that energy since conduction electrons adiabatically follow the acoustic wave acquiring its velocity 
$\sim 10^5cm/s$. 

As discussed in Sec.~\ref{intr}, a possible macroscopic effect (a charge separation resulting in local electric fields) of the slow acoustic pulses cannot lead to x-rays. Therefore the observed 
x-ray emission from shock wave exposed lead is paradoxical since it contradicts the known mechanisms.
\section{MATTER COLLAPSE}
\label{matt}
In our experiments we also studied another effect of shock waves, which reflect from a surface of the sandwich containing the lead foil of $0.06mm$ thickness. Each shock wave generates an acoustic
pulse propagating through the metal. At the energy used in our study, the lead foil could not be mechanically perforated by one shock wave. One shock wave only produced a residual plastic 
deformation of the foil which got slightly bent in the direction (the $x$ axis) of acoustic propagation. Further repeated shock waves lead to an accumulation of those deformations. After a
\begin{figure}[h!]
\centering
\includegraphics[width=5cm , bb=5 5 350 350]{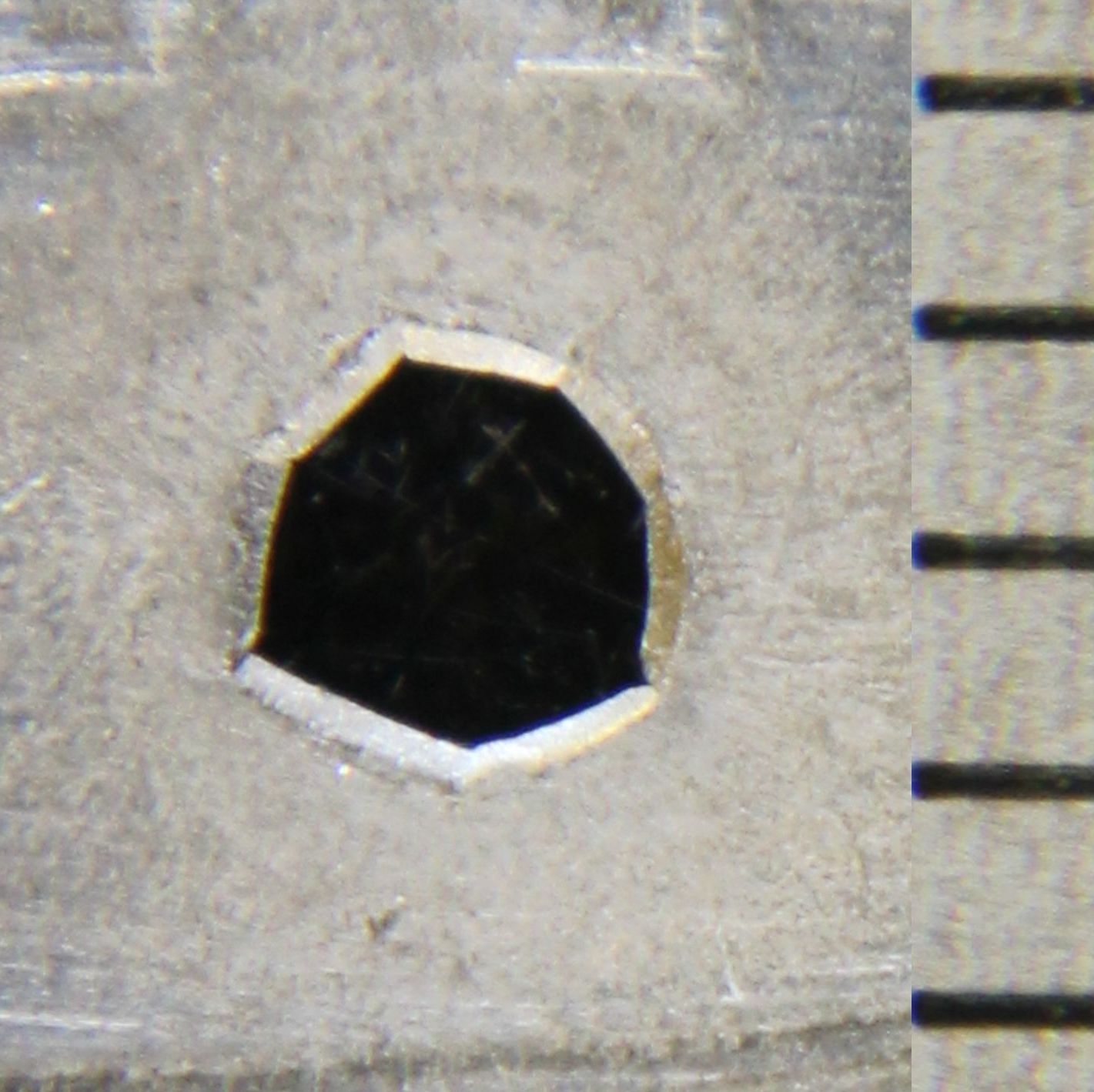}
\caption{\label{fig10}Amplified heptagonal hole of Fig.~\ref{fig8}. The scale is in millimeters.}
\end{figure}
\begin{figure}[h!]
\includegraphics[width=5cm , bb=5 5 350 350]{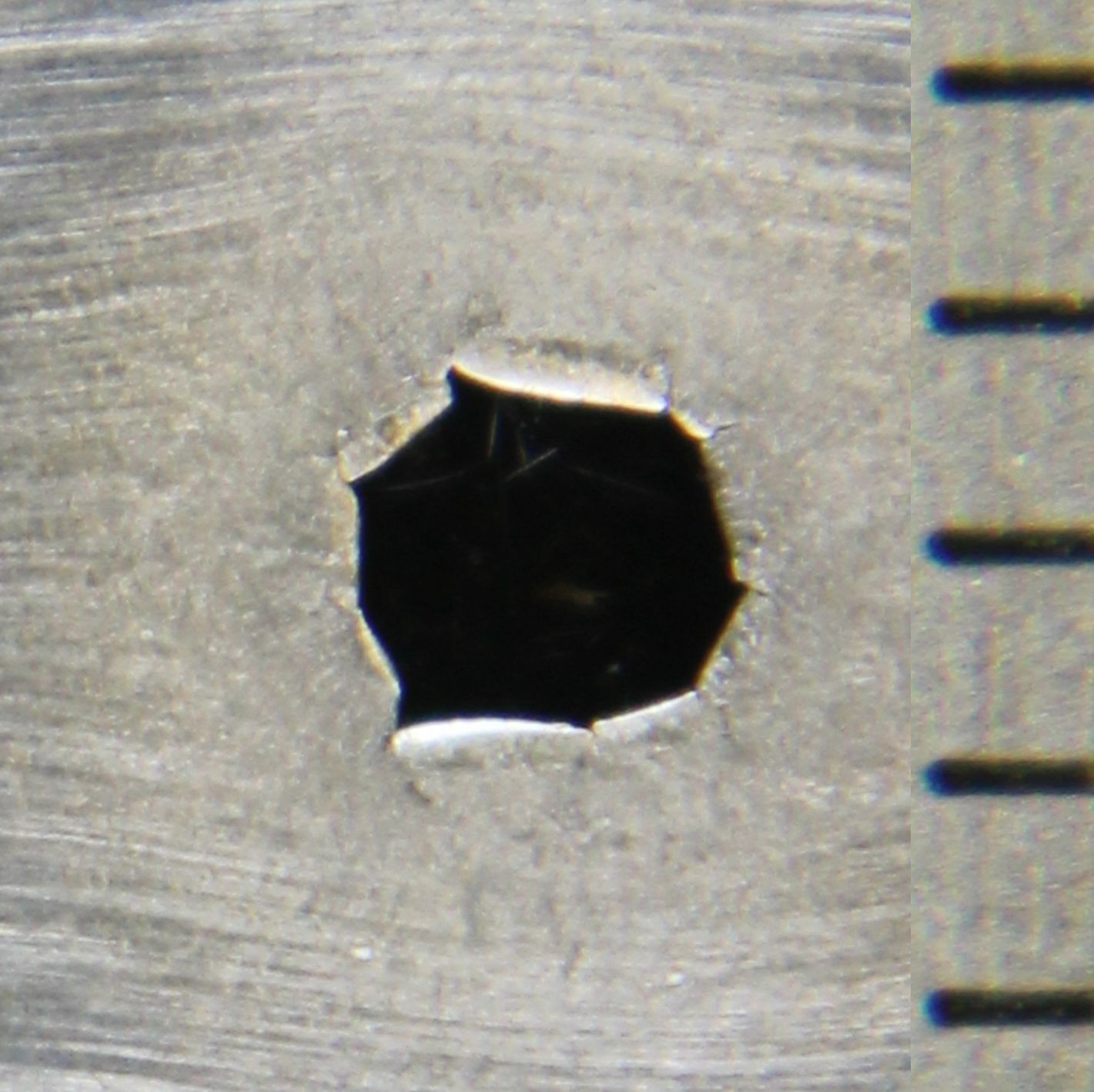}
\caption{\label{fig11}Octagonal hole in the lead foil after exposure to 500 shock waves with a peak positive pressure amplitude of $22MPa$. The shock wave velocity, calculated by Eq.~(\ref{0}), was 
$v\simeq 1.513\times 10^5cm/s$. The scale is in millimeters.}
\end{figure}
critical number (on the order of hundred) of shock waves the foil became perforated. This was accompanied by a strong deformation of the foil area surrounding the perforated region as in 
Fig.~\ref{fig7}. In this case the foil was in contact with water.

In Figs.~\ref{fig8} - \ref{fig11} the lead foil was hermetically sealed inside the factory plastic envelope (Fig.~\ref{fig2}) and was not directly exposed to underwater shock waves. However, those 
shocks, that weakly reflected off the water-plastic interface, also resulted in acoustic pulses in the metal.

Whereas the foil damages in Fig.~\ref{fig7} are due to mechanical actions of shock waves the damages in Figs.~\ref{fig8} - \ref{fig11} are substantially different. The scenario of pure mechanical 
creation of those holes by acoustic pulses would look straightforward. In this case, after accumulation of small deformations, the foil in the center would become thinner and finally get perforated. 
The lead matter, that initially filled out the hole, would be distributed around it. By the action of the contacting plastic film a rugosity of the foil would be created near the hole. 

\begin{figure}[h!]
\centering
\includegraphics[width=8.9cm , bb=10 10 500 355]{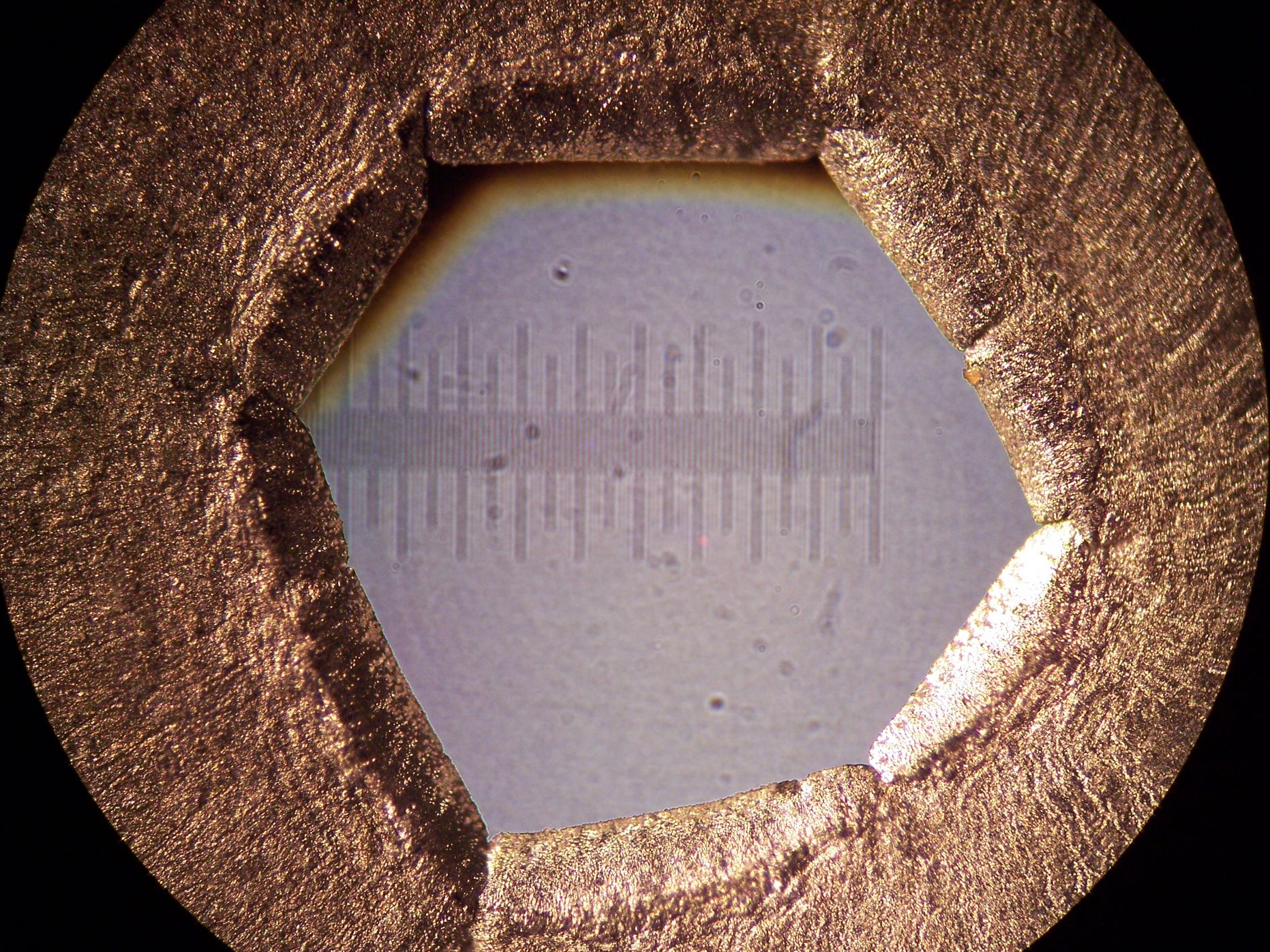}
\caption{\label{fig12}Microscope image of the lead foil surface, facing the incoming acoustic pulses, near the hole in Fig.~\ref{fig9}. The formation of micron size lead particles is visible. The 
distance between two large bars is $0.1mm$.}
\end{figure}

\begin{figure}[h!]
\centering
\includegraphics[width=8cm , bb=5 5 400 350]{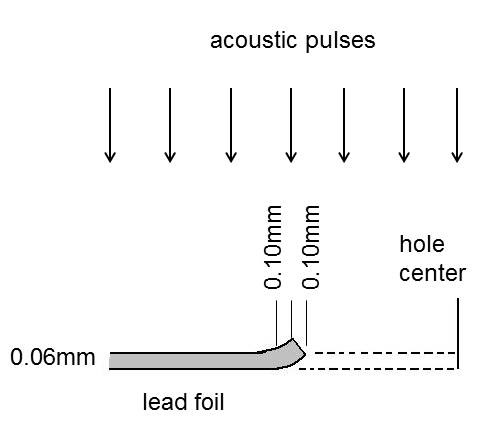}
\caption{\label{fig13}Schematic cross-section of the lead foil based on the microscope images. The chosen border point corresponds to the middle of the left-lower side of the hexagon in 
Fig.~\ref{fig12}. The foil was bent towards the incoming pulses.}
\end{figure}

\begin{figure}[h!]
\centering
\includegraphics[width=8.5cm , bb=5 5 400 300]{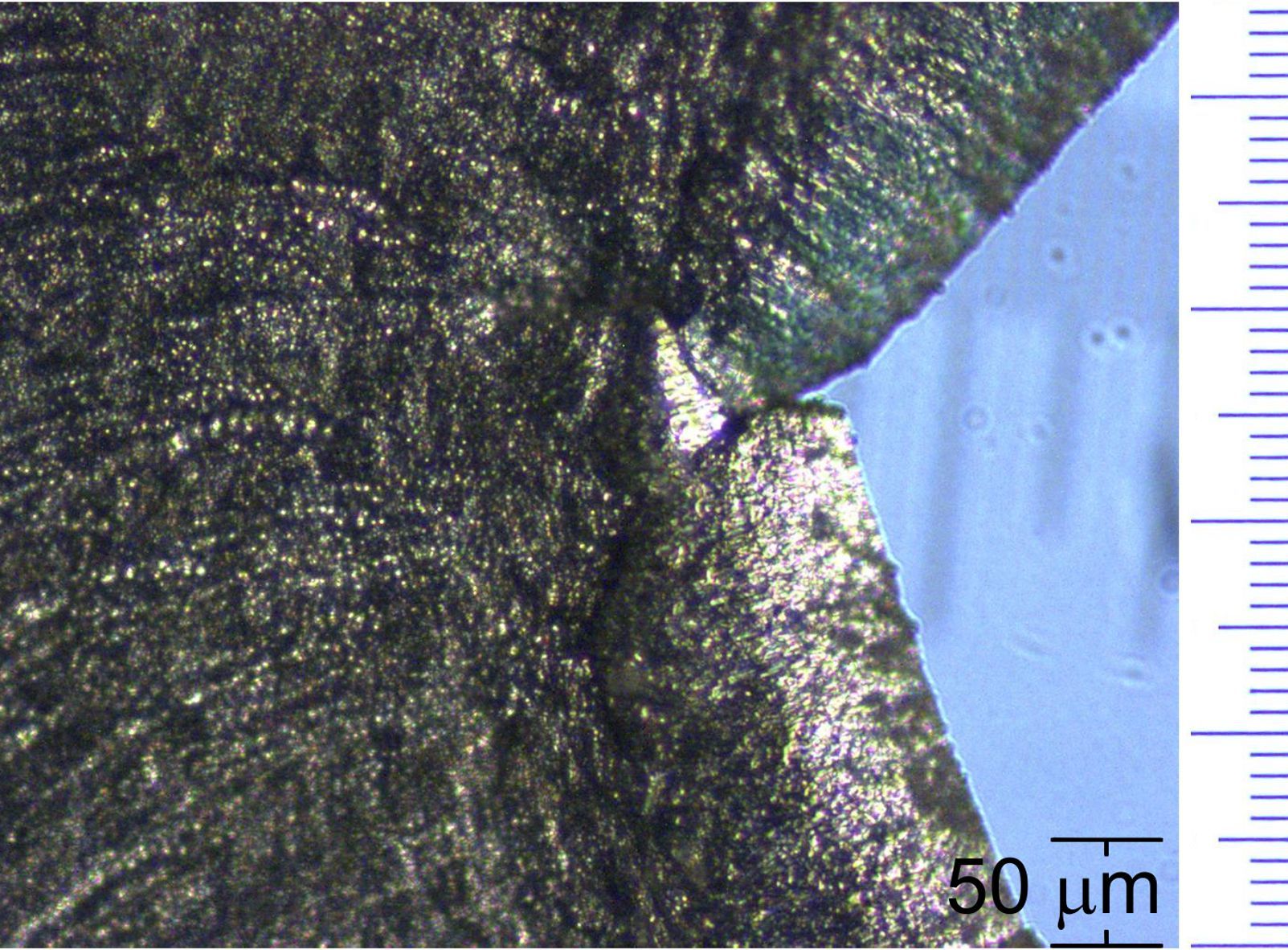}
\caption{\label{fig14}Microscope image of a part of the hole border. The total hole is shown in Figs.~\ref{fig9} and \ref{fig12}. The formation of micron-sized lead particles is visible.}
\end{figure}

The absence of a rugosity around the holes in Figs.~\ref{fig8} - \ref{fig11} contradicts the scenario of pure mechanical breaking. The microscope magnifications of the hole, shown in 
Figs.~\ref{fig12} and \ref{fig14}, are incompatible with a usual mechanical perforation. In Fig.~\ref{fig13} the cross-section of the foil at the hole shows that the foil was bent towards the 
direction of the incoming acoustic pulses. This also stands against a mechanical hole formation. Fig.~\ref{fig13} is schematically based on confocal  microscope analysis (not detailed here). According
to it, $25\%$ of the hole interior is collected on the hole border. This is also visible in Figs.~\ref{fig12} and \ref{fig14}. Therefore $75\%$ of the hole interior was somehow transferred away from 
the hole.

All holes, in Figs.~ \ref{fig8} - \ref{fig11}, look as if the missing lead matter would have been delicately removed with no disturbance of the surroundings. Those holes have a polygonal shape. The 
number of sides of the polygon increases from Fig.~\ref{fig9} (hexagon) to Fig.~\ref{fig11} (octagon) under the enhancement of the total acoustic flux to the foil. As follows from Figs.~\ref{fig12} 
and \ref{fig14}, micron-sized lead particles are formed on the lead foil surface facing the acoustic pulses. The particles are better seen on the microscope magnification of the part of the hole
border shown in Fig.~\ref{fig14}. Those particles are absent on the back side (exit side of the acoustic pulses) of the lead foil. 

There was another exciting phenomenon. The lead matter, looking as delicately removed from the holes, ``disappeared''. This corresponds to the aforementioned $75\%$ of the hole interior. It was not 
found in the envelope as lost macroscopic fragments. The plastic surfaces, which were in contact with the lead, looked as before shock wave action, i.e., they were not damaged, appeared to be clean, 
and without traces of lead. Fig.~\ref{fig10} corresponds to ``disappearance'' of $55mg$ of lead. Lead melting is impossible since a conversion of the entire acoustic energy into heat would increase 
the temperature of the missing lead part by $0.01K$ only. 

A conversion of the missing lead matter into the micron-sized particles, shown in Figs.~\ref{fig12} and \ref{fig14}, seems impossible since their total volume is approximately $1\%$ of the 
``disappeared'' hole interior.  

One can suppose that the missing $75\%$ of the hole interior was transferred by secondary acoustic pulses propagating along the lead foil away from the hole. In principle, such a pulse can be 
generated by the incident acoustic pulse, propagating across the lead foil, which has the duration $10^{-10}s$ (\ref{1}). So the hypothetical secondary pulse would have the spatial extension 
$10^{-10}s\times 10^{5}cm/s\sim 10^{-5}cm$ along the lead foil. It is easy to evaluate that each of 250 secondary pulses (referred to the hole in Fig.~\ref{fig9}) would carry away from the hole ten 
times increased lead density localized within $10^{-5}cm$. Therefore, the hypothetical secondary pulse would be a strong shock wave propagating along the lead foil. But there is no source for such
grandiose effect since the incident pulse, propagating across the lead foil, is weaker than a shock wave (Sec.~\ref{mech}). So secondary pulses cannot be generated and therefore do not transfer lead 
matter away from the hole.
  
The above phenomenon can be qualified as matter collapse, i.e., a non-conventional matter redistribution. Into what does the missing lead matter go over? 

\section{THE MECHANISM}
\label{mech}
The velocity of the shock waves in water in the focal plane in Fig.~\ref{fig1} is approximately $v\simeq 1.5\times 10^5cm/s$. It slightly exceeds the speed of sound in water $s_0$. In the 
macroscopic description there is a jump of velocity at the shock wave front. In reality this jump is smeared out over approximately $\Delta x\sim 0.15\mu m$. Therefore the characteristic time, 
related to the shock wave front propagation, is given by Eq.~(\ref{1}).

The metal foil was placed at the focal region of the shock wave source with the size of a few millimeters. Within this area one can consider a plane shock wave which collides the foil normally (in 
the $x$ direction). We are interested on short time processes occurring during the front wave time (\ref{1}). The density $\rho$ of the metal is larger than the density $\rho_0\simeq 1g/cm^3$ of 
water. Therefore the shock front in water reflected almost elastically from the metal border. In this process the momentum $2(\rho_0\Delta x)v$ per unit area of the water-metal interface was 
transferred to the metal. See also \cite{BEN, BEI}. This momentum, acquired by the metal, leads to the acoustic pulse $u(x-st)$ propagating from its surface. Here $s$ is the speed of sound in the 
metal and $u(x)$ is the positive longitudinal displacement localized on $x\sim s\Delta t$. It results in the local variable density $(1-\partial u/\partial x)\rho$ in the metal, where $\rho$ is the 
equilibrium density. The local velocity in the metal is $V=-s\,\partial u/\partial x$. 

Below we estimate the balance of momentum transfer per unit area of the water-metal interface. It can be written as 
\begin{equation}
\label{2}
\Big\langle\left(s\Delta t\right)\left(1-\frac{\partial u}{\partial x}\right)\rho V\Big\rangle =2\left(v\Delta t\right)\rho_0 v,
\end{equation}
where $s\Delta t$ and $v\Delta t$ are the lengths of the pulse (along the $x$ axis) in metal and water, respectively. Angular brackets mean an average on $x$. Since the average of 
$\partial u/\partial x$ is zero, one should account the square of this term in the left-hand side of (\ref{2}). The maximal velocity $V$ in the pulse, propagating in the metal, is estimated as
\begin{equation}
\label{3}
V\simeq v\,\sqrt{\frac{2\rho_0}{\rho}}\,.
\end{equation}
For lead $\rho\simeq 11.3g/cm^3$, $s\simeq 1.26\times 10^5cm/s$, and therefore the maximal value of $V/s=-\partial u/\partial x$ can be estimated as 0.49. This estimate is approximate since we use 
the limit of small $\partial u/\partial x$. However it shows that $\partial u/\partial x\sim 1$ and the pulse, propagating through the lead, is not a shock wave but rather a strong acoustic 
perturbation. 

The two paradoxical phenomena, the x-ray emission and matter collapse, are expected to be consequences of a certain mechanism which underlies both. To understand the basis of these intriguing 
phenomena let us consider the concept of anomalous states developed in Ref.~\cite{IVLEV}.

In condensed matter, certain macroscopic perturbations can form anomalous electron wells due to a local reduction of zero point electromagnetic energy. These wells are narrow, on the order of the 
Lamb radius $r_L\simeq 0.82\times 10^{-11}cm$ \cite{IVLEV}, and are localized around an atomic nucleus which is of $10^{-13}cm$ size. The well depth is $\sim \hbar c/r_L\simeq 2.4MeV$. The electron 
spectrum in the well is continuous and non-decaying \cite{IVLEV}. The latter is true when the well with electrons is at rest. In a condensed matter, under thermal vibrations resulting in 
Bremsstrahlung of the captured electrons, states in the anomalous well acquire a small but finite width. That well with captured electrons may be treated as an anomalous atom. The wells can be 
created by some perturbation which rapidly varies in space, on the scale $10^{-11}cm$. 

In condensed matter such perturbations may relate to acoustic pulses. In this process the short scale is the length of the standing de Broglie wave of reflected lattice atoms resulting in 
a spatial variation of charge density. In the acoustic pulse, propagating through the metal, the lattice site acquires the velocity $V$ (\ref{3}). When the pulse continues its motion, the site 
returns to its initial position with the velocity $-V$ due to a reflection from other sites. In this process the quantum interference of forth and back motions results in the modulation of the 
charge density on the scale
\begin{equation}
\label{4}
\Delta R=\frac{\hbar}{2MV}\,,
\end{equation}
where $M$ is the mass of the lattice site. For lead $M\simeq 3.44\times 10^{-22}g$ and the above estimates give $\Delta R\simeq 2.99r_L$. Therefore the perturbation of lead by the acoustic pulses 
is effective for creation of anomalous electron states. 

The quantum coherence of the incident and reflected de Broglie waves can exist if it is not destroyed by thermal fluctuations. For this reason, the velocity of the macroscopic motion $V$ of 
lattice sites should exceed their velocity $V_T=\sqrt{T/M}$ of thermal motion. For lead at room temperature ($20^o C$) the velocity $V$ is approximately $6.1V_T$.  

One should note that in copper the above mentioned acoustic pulses will hardly create anomalous states. The copper density $\rho=8.96g/cm^3$ and the mass of the copper atom $M=1.055\times 10^{-22}g$ 
correspond to $\Delta R\simeq 8.4r_L$ which is a too long spatial perturbation compared to $r_L$. So, heavy metals are preferable for formation of anomalous states. Note that for copper at room 
temperature ($20^o C$) the velocity $V$ is approximately $3.6V_T$.  

According to the Thomas-Fermi approach \cite{LANDAU1}, in heavy atoms the energy to remove all electron from a neutral atom is $\Delta E\simeq 16Z^{7/3}eV$. For lead (atomic number $Z=82$) 
$\Delta E\simeq 5.7ZkeV$. Since the Thomas-Fermi method is approximate, in reality one can estimate the ground state energy per electron in lead as minus a few $keV$. 

After the fast (within the time $\hbar/1MeV\sim 10^{-21}s$) formation of the anomalous well the initial atomic state, in the $keV$ region, is ``unable'' to immediately vary and remains the same as 
before. But now this state is non-stationary corresponding to an electron flow toward the anomalous well to form the anomalous atom. A typical time of this process is $\hbar/1keV$. Faster processes 
are less effective due to strong spatial oscillations (compared to the initial wave function) of new states localized in the anomalous well. This is similar to smallness of a matrix element when 
one wave function rapidly oscillates. So, according to quantum mechanical uncertainty, the formation of the anomalous well is accompanied by quanta emission in the $keV$ region. Further $MeV$ quanta 
emission from the formed well is weak since the states in the well are almost non-decaying. An intensive high energy ($MeV$) emission can be induced by some external sources. 

One can see that the deep (of the $MeV$ range) anomalous well, from where the $keV$ x-ray emission occurs, is formed by adiabatic acoustic perturbations with the typical time 
$\Delta t\sim 10^{-10}s$. This resolves the paradoxical observation of x-ray emission under the slow varying driving force (Sec.~\ref{xray}). 

There is another unusual aspect of anomalous atoms. The size of this atom ($\sim 10^{-11}cm$) is three orders of magnitude smaller than the conventional one. If in a part of a solid all atoms 
undergo the transition to the anomalous state that macroscopic region reduces $10^9$ times in volume. This process can be qualified as matter collapse. The collapsed matter (with $10^9$ times
enhancement of density) looks as a dramatically different concept \cite{IVLEV}. 

In our experiments the missing macroscopic part of lead, initially filling the areas of the polygonal holes in Figs.~\ref{fig8} - \ref{fig11}, did not disappear but could be converted into collapsed 
matter of micron size. Such ``speck of dust'', invisible even with a magnifying lens, imitates matter ``disappearance''. Another possibility is that the observed micron-sized particles on the lead 
surface are collapsed (or partly collapsed) matter. Further research is required to establish which scenario is realized. Anyway, it would be extremely unusual to mechanically collect after an 
experiment that heavy matter artificially created. That heavy matter, up to $10^9$ times enhancement in density, is unknown to be found in nature. 

The paradoxical x-ray emission and the lead matter ``disappearance'' cannot be explained by a combination of known effects. A different concept is required which underlies both intriguing
phenomena. The concept of anomalous states satisfies this condition. It explains the common mechanism of both phenomena and also why they are observed in lead but not in copper. 

\section{DISCUSSION}
\label{disc}
The main purpose of this paper is to introduce the unexpected paradoxical phenomena providing convincing experimental arguments for their existence. Further quantitative studies, for example 
spectrum measurements of the emitted x-rays, are to be done. The second purpose is to link those paradoxical phenomena to the mechanism likely responsible for them. 

As known, usual applications of shock waves relate to their mechanical actions on matter. We report a completely different type of a shock wave action. Two intriguing phenomena were observed. 

1. {\it Sound into x-rays}. A strong x-ray emission from the lead under the action of acoustic pulses which are extremely adiabatic, $10^8$ times slower, compared to fast processes of 
x-ray generation was observed. The estimated x-ray flux, during acoustic pulse action, was approximately $10^7$ times higher than the flux from a dental x-ray tube. Our phenomenon strongly differs
from ones when x-ray emission occurs due to known mechanisms initiated by extremely strong shock waves (artificial explosions, processes in supernovae in astrophysics, etc.). 

2. {\it Matter collapse}. ``Disappearance'' of a macroscopic part of lead matter exposed to acoustic pulses was noticed. The ``disappeared'' matter has the shape of a polygonal hole in the lead 
foil with non-damaged surroundings of the hole. The polygonal shape of spots and holes confirms a resemblance between two phenomena. A direct mechanical effect of the acoustic pulses, as a cause of 
that hole formation, is excluded. In that case, the acoustic pulses would strongly damage the hole surroundings as was observed in previous experiments.  

The two above paradoxical phenomena were discovered in table-top experiments and remind, at first sight, routine acoustic studies. This impression is not correct. Both phenomena cannot be explained 
by a combination of known effects. It turns out that a fundamentally new mechanism has to underlie them. The concept of electron anomalous states, which encouraged the experiments and specified 
main features of them, likely is this mechanism. 

It relates to the formation in condensed matter of anomalous electron wells, because of a local reduction of zero point electromagnetic energy. The wells are narrow, $10^{-11}cm$, and deep, on the 
order of a few $MeV$. The well formation is due to short scale ($10^{-11}cm$) modulations of charge density caused by the interference of de Broglie waves of lattice sites. This interference occurs, 
on that short scale, between the incident and reflected lattice site participating in shock wave motion with the velocity $V$. The thermal motion of lattice sites, with the velocity $V_T$, cannot 
destroy that interference since in lead $V_T$ is substantially smaller than $V$.

Atomic electrons are collected inside the anomalous well forming an anomalous atom and resulting in two effects. First, they emit x-ray quanta by transitions to lower levels. Second, the anomalous 
atom is three orders of magnitude smaller than a usual one. The lead matter, that initially filled the hole region, transformed into anomalous state reducing its macroscopic volume $10^9$ times. In 
other words, the macroscopic volume is converted into a heavy particle of micron size. This ``speck of dust'', invisible even with a magnifying lens, imitates matter ``disappearance''. The origin 
of micron size lead particles on the lead surface, facing acoustic pulses, may be of the above mentioned nature but this needs further studies.

The theory of anomalous states \cite{IVLEV} explains (i) the paradoxical x-ray emission caused by extremely adiabatic perturbations, (ii) paradoxical matter collapse, and (iii) the lack of effect in 
copper in contrast to lead. A detailed study of the phenomena discussed is not the goal of this paper. 

Furthermore, there is another aspect related to quanta emission from anomalous matter. All electrons of the single anomalous lead atom, in principle, can emit approximately $250MeV$ by transitions 
to lower states \cite{IVLEV}. These processes do not occur automatically since anomalous states are weakly decaying. One can ask the question: what kind of external perturbation triggers off that 
avalanche releasing $250MeV$ per atom? In this case $36mg$ of lead would release $4.18\times 10^9J$ originating from a reduction of zero point electromagnetic energy \cite{IVLEV}.\\
~

~
\section{CONCLUSIONS}
\label{conc}
Acoustic pulses, propagating through lead, unexpectedly resulted in strong x-ray emission (sound into x-rays). Those pulses are extremely adiabatic compared to atomic processes of x-ray generation 
which have the formal probability $\exp(-10^8)$. Bremsstrahlung mechanisms are excluded.

The lead foil, exposed to acoustic pules, misses a part of its area in the shape of a polygonal hole of the size of $\sim 2mm$ (matter collapse). The missing polygon of the lead foil looks as a 
delicately removed part with no damage at the hole surroundings as it should be after a strong mechanical breaking. That missing polygonal lead matter seems ``disappeared'' because no traces of it 
were found. 

The discovered phenomena of {\it sound into x-rays} and {\it matter collapse} require a fundamentally new mechanism to underlie them. The concept of electron anomalous states \cite{IVLEV} is likely 
that mechanism.

\acknowledgments
The work was supported by CONACYT through grant 237439.

\end{document}